\documentclass[sn-nature]{sn-jnl}

\usepackage{graphicx}%
\usepackage{multirow}%
\usepackage{amsmath,amssymb,amsfonts}%
\usepackage{amsthm}%
\usepackage{mathrsfs}%
\usepackage[title]{appendix}%
\usepackage{xcolor}%
\usepackage{textcomp}%
\usepackage{manyfoot}%
\usepackage{booktabs}%
\usepackage{algorithm}%
\usepackage{algorithmicx}%
\usepackage{algpseudocode}%
\usepackage{listings}%
\usepackage{siunitx}
\usepackage{lineno}
\usepackage{hyperref}

\begin{document}

\title[Article Title]{Tailored 3D microphantoms: an essential tool for quantitative phase tomography analysis of organoids}

\author[1]{\fnm{Micha\l{}} \sur{Ziemczonok}}
\author[2]{\fnm{Sylvia} \sur{Desissaire}}
\author[3]{\fnm{J\'er\'emy} \sur{Neri}}
\author[1]{\fnm{Arkadiusz} \sur{Ku\'s}}
\author[2]{\fnm{Lionel} \sur{Herv\'e}}
\author[2]{\fnm{C\'ecile} \sur{Fiche}}
\author[2]{\fnm{Guillaume} \sur{Godefroy}}
\author[3]{\fnm{Marie} \sur{Fackeure}}
\author[3]{\fnm{Damien} \sur{Sery}}
\author[1]{\fnm{Wojciech} \sur{Krauze}}
\author*[3]{\fnm{Kiran} \sur{Padmanabhan}} \email{kiran.padmanabhan@ens-lyon.fr}
\author*[2]{\fnm{Chiara} \sur{Paviolo}} \email{chiara.paviolo@cea.fr}
\author*[1]{\fnm{Ma\l{}gorzata} \sur{Kujawi\'nska}}\email{malgorzata.kujawinska@pw.edu.pl}

\affil[1]{\orgname{Warsaw University of Technology, Institute of Micromechanics and Photonics}, \orgaddress{\street{Boboli 8 Street}, \city{Warsaw}, \postcode{02-525}, \country{Poland}}}

\affil[2]{\orgdiv{Univ. Grenoble Alpes}, \orgname{CEA, Leti}, \postcode{F-38000} \city{Grenoble}, \country{France}}

\affil[3]{\orgdiv{Institut de G\'enomique Fonctionnelle de Lyon}, \orgname{Ecole Normale Sup\'erieure de Lyon, CNRS}, \orgaddress{\street{Universit\'e Claude Bernard Lyon 1}, \city{Lyon}, \postcode{UMR 5242}, \country{France}}}

\abstract{We present a novel approach for benchmarking and validating quantitative phase tomography (QPT) systems using three-dimensional microphantoms. These microphantoms, crafted from biological and imaging data, replicate the optical and structural properties of multicellular biological samples. 
Their fabrication featuring refractive index modulation at sub-micrometer details is enabled by two-photon polymerization. We showcase the effectiveness of our technique via a round-robin test of healthy and tumoral liver organoid phantoms across three different QPT systems. This test reveals sample- and system-dependent errors in measuring dry mass and morphology. Tailored microphantoms establish the gold standard for the optimization of hardware setups, reconstruction strategies and error assessment, paving the way for novel non-invasive, label-free measurements of 3D biological samples.}

\maketitle

\section*{Significance statement} 
Organoids are multicellular biological structures that begin to show organ level functions, but are challenging to monitor and measure at the cellular level. Quantitative phase tomography has emerged as novel imaging technique to meet this demand, capable of imaging such specimens non-destructively and in three dimensions. We propose a novel methodology for the design and fabrication of microphantoms that perfectly replicate optical and structural properties of cultured organoids. These phantoms are crafted from biological and imaging data to mimic refractive index modulations of real organoids and can be tailored for a particular biological imaging task. We believe that presented phantoms will become an essential tool for optimization and certification of new instruments for reliable study of organoids.

\section{Introduction}\label{sec:intro}
Over the past decade, organoid models have emerged as a powerful platform to recapitulate animal physiology \emph{ex vivo}. Organoids are complex three-dimensional (3D) multicellular structures derived from stem cells present in tissues or from induced pluripotent stem cells that grow, divide and self-organize \cite{zhao_organoids_2022}. These models represent a key advance on conventional \emph{in vitro} 2D cell cultures, as they effectively  capture the complex physiology of multicellular systems and allow for modeling pathological events such as tumorigenesis in physiologically-relevant environments  \cite{Duval2017_2Dvs3D_cell_culture}. Optically, organoids require imaging and monitoring at scales spanning from \SI{100}{nano meter} (subcellular features) to mm (fully-grown). They are also dense multi-scattering structures, which limit their imaging and analysis without optical clearing \cite{susaki_perspective_2021}. While the most common imaging technique for 3D characterization relies on fluorescent staining, the use of exogenous tags has intrinsic disadvantages including photobleaching and phototoxicity over long-term time-lapse imaging \cite{Pawley2005}. In recent years label-free quantitative phase tomography (QPT) methods have been increasingly used in characterizing biological systems such as 2D cell cultures, 3D organoids and spheroids \cite{Astratov2023_roadmap_on_labelfree, nguyen_quantitative_2022, park_quantitative_2018}. QPT retrieves the refractive index (RI) of the sample relative to the surrounding media within each imaging voxel, resulting in a quantitative 3D refractive index distribution within the sample. 

In order to tackle imaging of more and more complex biological specimens, the advances in QPT focus on more accurate measurements of the complex field disturbed by the object, synthesizing 3D aperture with more and more angular and spectral components \cite{Verrier2024_ODT_review} and utilizing new forward models \cite{lim2019high, chen2020multi, pham2020three}. In principle, QPT provides subcellular resolution typically down to 150 nm half-pitch and the accuracy and precision of RI enables segmentation of cellular compartments such as the nucleus, nucleoli or lipid droplets. However, the technique becomes less feasible when approaching diffusion imaging regime, mainly due to scattering of thick samples (\SI{>100}{\micro \meter}) that exhibit high RI gradients. In order to push the limits of QPT applications in such scattering samples, longer illumination wavelengths can be used \cite{Ossowski2022} or the coherence of the light can be decreased (e.g. light-emitting diode array \cite{Lee2023, chowdhury_high-resolution_2019, tian_3d_2015}).

While QPT has the capacity to visualize and quantify sample features in 3D, further analyses of the accuracy of the reconstructed RIs and the quality assessment metrics are required. Recently, we proposed a 3D-printed scattering microphantom with known geometry and RI distribution \cite{krauze20223d}. While microphantoms enable the quantification of RI errors in 3D reconstructions, these artificially designed targets still lack the typical biological diversity and RI entropy seen in complex biological models, such as 3D organoid cultures. 
\begin{figure}[b]
    \centering
    \includegraphics[width=0.9\textwidth]{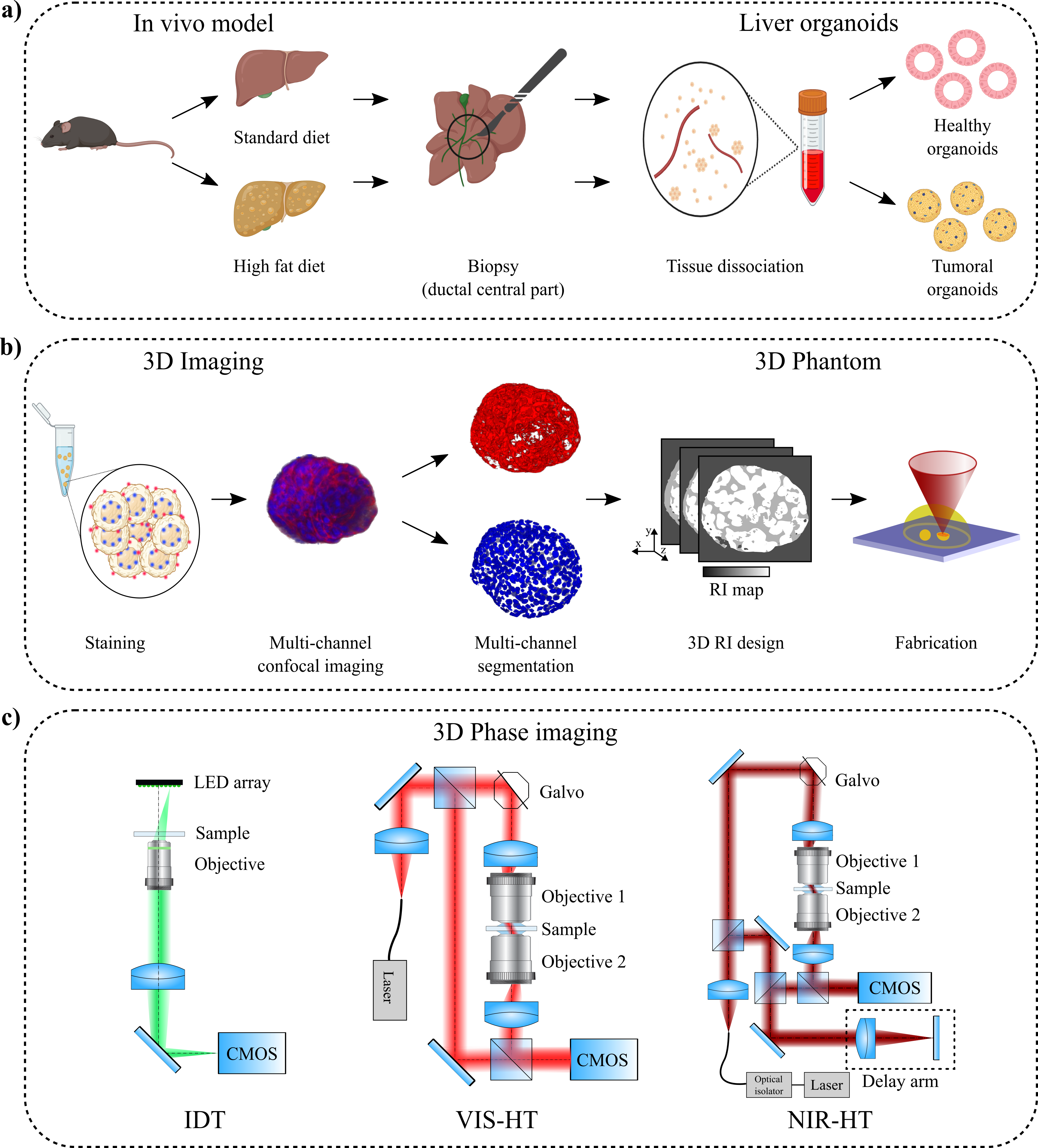} 
    \caption{Overview of the methodology for the design and fabrication of the 3D microphantoms. a) Preparation of targeted imaging specimen, in this case healthy and tumoral murine liver organoids. b) Imaging with a suitable technique to reveal structural and/or functional properties of the sample (e.g. confocal imaging), used for inferring the 3D refractive index distribution of the fabricated phantom. c) Validation and benchmarking of different QPT imaging systems, namely intensity diffraction tomography (IDT) and two holographic tomography systems working in the visible (VIS-HT) or near-infrared spectral range (NIR-HT).}
    \label{fig-overview}
\end{figure}
Here we propose a novel methodology to create 3D-printed micro-phantoms that mimic the morphological details of cultured 3D organoids for the purpose of benchmarking and validating 3D QPT systems. The pipeline begins with the characterization of liver organoids, generated from mouse models for liver cancer that include groups of animals that were raised under either healthy or modified high-fat diets that promoted liver pathologies (Fig.~\ref{fig-overview}a). The structural and functional properties of the organoids were retrieved with fluorescently-labelled confocal microscopy and then converted into RI values. The phantoms were fabricated via 2-photon polymerization, capable of locally tuning the RI contrasts to mimic different biological structures and measurement conditions (Fig.~\ref{fig-overview}b). The obtained phase-calibrated targets were then used for characterization of different QPT imaging setups devoted to imaging of 3D biological samples (Fig.~\ref{fig-overview}c). This processing chain establishes the gold standard for characterization of QPT imaging, paving the way for novel targets that can be designed \emph{ad hoc} for different biological experimental models and medical diagnosis applications.

\section{Results}\label{sec:results}
\subsection{3D culture of liver organoids}\label{sec:organoid_culture}
Hepatic organoids are used to study the function of the liver \emph{ex vivo} and offer the potential towards applications in the clinical setting. To generate wildtype or tumoral murine liver organoids, liver stem cells and ductal structures from healthy mice or littermates that were raised on a choline-deficient high-fat diet (CDAHFD) for 42 weeks were isolated and cultured in 3D matrigel (Fig.~\ref{fig-overview}a). The CDAHFD regime has been shown to promote liver tumorigenesis in otherwise healthy animals \emph{in vivo} \cite{Ikawa-Yoshida2017}. After establishment of 3D cultures, healthy mouse liver organoids (MLO) or tumoral liver organoids (MTLO) were analyzed by immunofluorescence for cytoskeletal (phalloidin, which stains the actin cytoskeleton) and epigenetic markers (H2A.Z protein) to confirm their pathological relevance using confocal microscopy. Unlike the MLOs which were characterized as a monolayer of cells organized around an internal cavity, MTLOs were typically smaller with cells invading the lumen. While MLOs showed typical basal staining of the actin cytoskeleton (Phalloidin signal, red fluorescence channel) as well as histone H2A.Z expression (green fluorescence channel), MTLOs displayed reorganization of the actin cytoskeleton as well as intense H2A.Z signals reflecting observations from human hepatocellular carcinoma patient liver biopsies \cite{Tang2020}. 
\begin{figure}[t]
    \centering
    \includegraphics[width=1\textwidth]{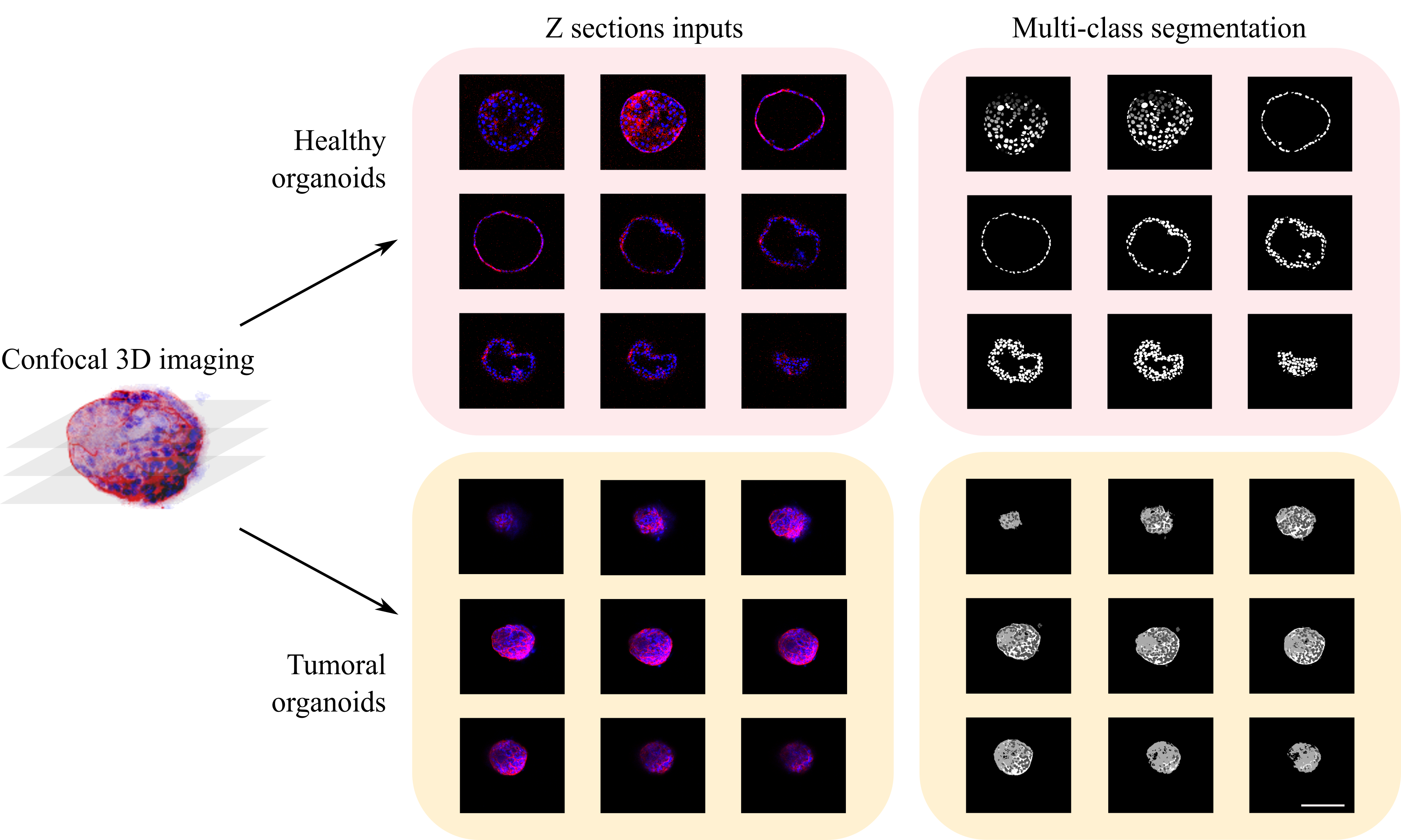} 
    \caption{Selected cross-sections of fluorescence confocal imaging and corresponding segmentations for the healthy and tumoral organoid. Cells nuclei are shown in blue (Hoechst dye), while actin in red. The fluorescence datasets were then segmented in CellPose using different intensity thresholds. }
    \label{fig:segmentation}
\end{figure}

3D fluorescent images of either MLO or MTLO were split into individual channels and directly segmented using CellPose (Fig.~\ref{fig-overview}b). This approach takes advantage of the different fluorescent channels to detect cellular contours or more specific intracellular organelles. This concept of functional imaging guiding the phantom design can be easily adapted to various imaging techniques marking specific biological structures (e.g. confocal Raman spectral imaging \cite{lalone_quantitative_2023}). Figure \ref{fig:segmentation} shows different confocal \emph{z} sections and corresponding segmentations for healthy or tumoral organoids. The staining specifically marked nuclei and actin in order to showcase the characteristic morphological features of liver organoids at different stages of disease (i.e. presence of the internal lumen).

\subsection{Generation of 3D-printed organoid phantoms}\label{sec:3dprint}
\begin{figure}[htbp]
    \centering
    \includegraphics[width=.9\textwidth]{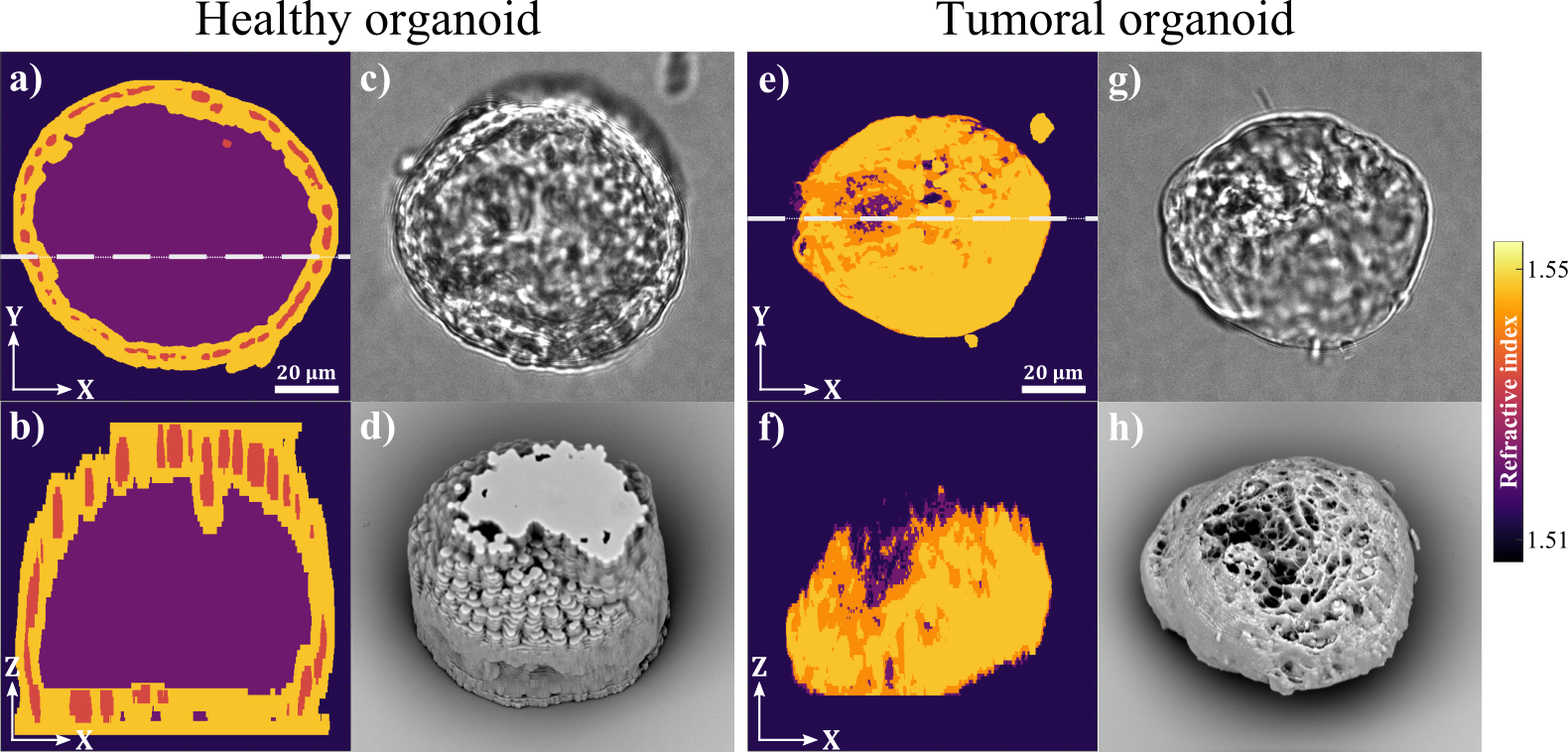} 
    \caption{The design and fabrication preview of the organoid phantoms: a-b) 3D RI cross-sections, c) bright-field image and d) SEM image of the healthy MLO. Panels e-h) show corresponding images for the MTLO phantom.}
    \label{fig:design}
\end{figure}

The volumes segmented based on fluorescence data were converted to the RI values considering the physiological 3D RI distribution and the range of values accessible in the photosensitive resin during fabrication. Phantom fabrication was performed using two-photon polymerization (TPP) lithography, which is capable of producing 3D polymeric structures with locally-adjustable RIs. Figure \ref{fig:design} shows the designed 3D RI cross-sections and post-fabrication scanning electron microscopy images of resulting healthy and tumoral organoid phantoms. 
Nuclei and cytoskeleton were assigned the RI contrast of 0.015 based on values established in literature \cite{Liu2016_cellRI, Gul2021_cellRI, Nguyen2022_cellRI}, while the maximal contrast in the sample, that is between the phantom and the surrounding medium, was set to 0.0374 in order to mimic the conditions of organoid growth in matrigel (see Materials and Methods).

MTLOs differ from the MLOs substantially, in that they were smaller and most notably characterized by the absence of an internal lumen and reorganization of the actin cytoskeleton
(Fig.~\ref{fig:design}e-h). The RI distribution was therefore derived from the actin fluorescence channel and assigned three distinct values based on the fluorescence intensity. This leads to a dense, disordered phantom full of crevices and debris that is a representative case for 3D RI of the MTLO. The internal RI contrast between the three labels was in the range of 0.015, while the maximal contrast between the phantom and the medium is 0.0374, as in the case of healthy MLO.

One of the main challenges in 3D organoid imaging stem from their size and RI contrast resulting in multiple scattering of visible light, thus positioning this kind of experiments somewhere between single-scattering and diffusive imaging regimes \cite{Yoon2020_scattering_review,Lambrou2021_scattering_review}. To meet this challenge, new approaches in instrumentation and numerical tools are being developed, such as coherence gating or deep learning, to limit or account for multiple scattering \cite{Astratov2023_roadmap_on_labelfree}. Biologically, optical clearing methods can also be used to normalize the RIs throughout the sample bringing it into the single-scattering mode. This procedure can produce high-quality 3D images but alters the natural 3D molecular distribution of living samples \cite{Boothe2017_RImatching_medium,Lee2022_immersionRImatching}. 
To match these imaging conditions with out phantoms, we created additional pair of organoids mimicking the optically-cleared samples, in which we decreased the maximal RI contrast from 0.0374 down to 0.0061 by partial polymerization of the surrounding medium. Thus, this approach to the fabrication enables evaluation of the systems for the weakly-scattering samples. Another use case for such low-RI contrast sample is that it can be used for validation of the fabrication process and especially intricate internal details of the phantoms, since the sample in measured under much more favourable imaging conditions that better comply with approximations embedded into the tomographic solvers.
\subsection{Quantitative phase imaging}\label{sec:qpi_measurement}
The 3D-printed phantoms were measured using three different QPT systems to verify their RI quantification capabilities and applicability for imaging 3D multi-scattering biological samples (Fig.~\ref{fig-overview}c): \emph{i}) an intensity diffraction tomograph (IDT), \emph{ii}) a holographic tomograph working in the visible spectral range (VIS-HT), and \emph{iii}) a holographic tomograph working in the near-infrared range (NIR-HT). These configurations spanned different instrumental and numerical approaches, including wavelength and coherence of the illumination source, type and amount of data captured, processing frameworks and tomographic reconstruction algorithms.
Since each system has been optimized for their respective hardware-software configuration, the phantoms were treated as an unknown "round-robin" artifact and the results were analyzed in end-to-end fashion. Detailed description of the hardware and reconstruction software can be found in Materials and Methods, while raw measurement data and the reconstructed 3D RI are publicly available (see Data Availability).

The IDT microscope uses an LED array to acquire intensity-only images of the sample at different illumination angles. The sample’s optical properties (RI and absorption) are reconstructed using the beam propagation method embedded inside a deep learning framework, where the layers of the algorithms encode the optical properties of the sample. The microscope has been designed to image large specimens 
$>$\qtyproduct[product-units = single]{100 x 100 x 100}{\cubic\micro\meter} with subcellular resolution in a 4D configuration (3D+time) \cite{Pierre:22}. The two HT systems capture off-axis holograms resulting from interference of the object and reference beam in order to retrieve complex amplitudes of the field in the sample plane. The amplitude and phase of each projection are used to retrieve the 3D RI via Fourier diffraction theorem. The illumination direction is driven by the dual-axis mirror, which combined with the high-numerical aperture of both condenser and imaging lens enables high illumination angles of up to \SI{45}{\degree}. 

\begin{figure}[htbp]
    \centering
    \includegraphics[width=1\textwidth]{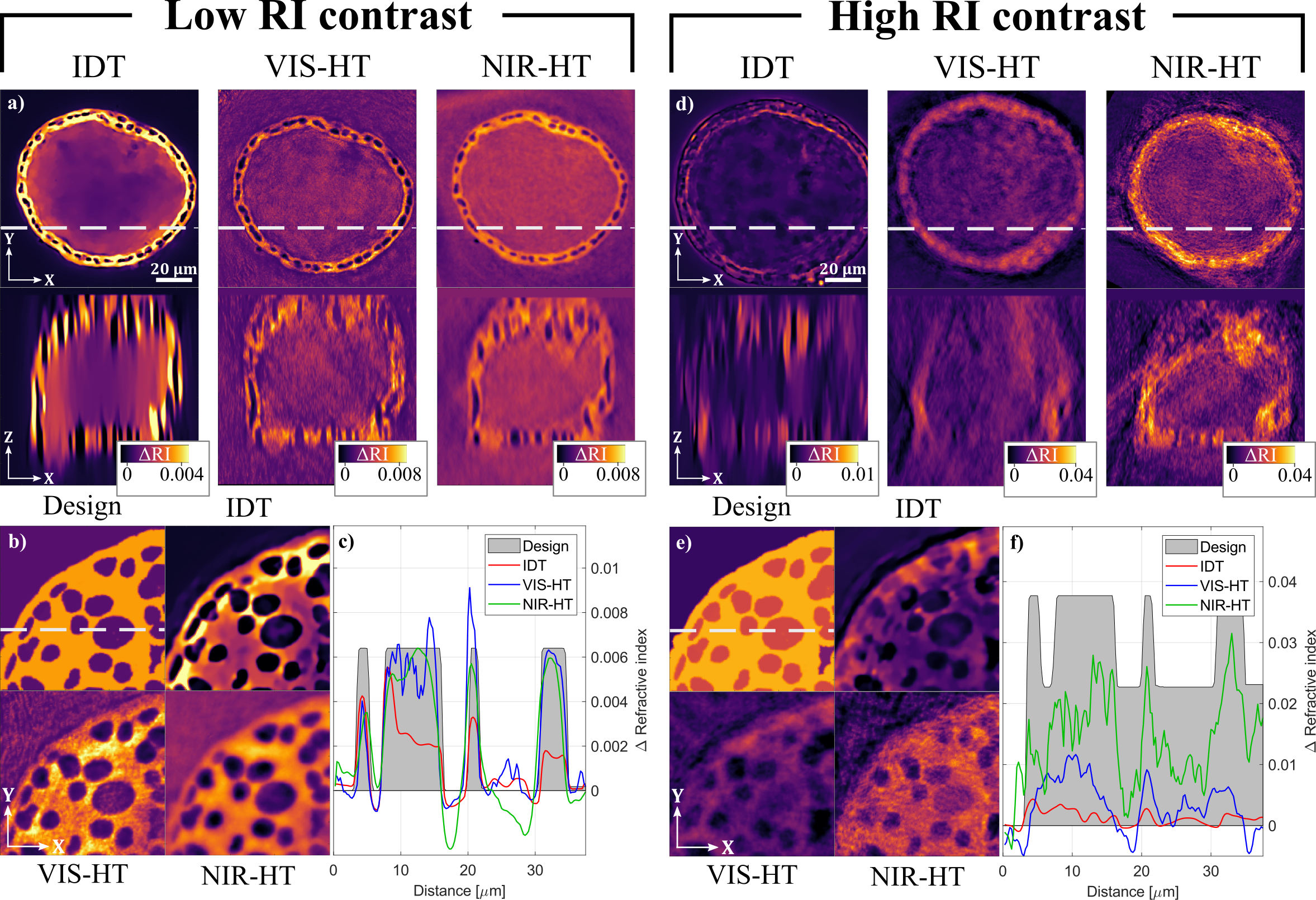} 
    \caption{3D RI reconstructions of a healthy MLO phantom measured under low (a-c) and high (d-f) RI contrast. Panels a) and d) present lateral and vertical cross-sections for IDT, VIS-HT and NIR-HT systems, while the b) and e) show a region of interest near the base of the organoid and the RI profile (c, f) across the dashed line in b) and e). Note that the RI colorbar for the IDT system is adjusted for visual contrast and is different from the HTs.}
    \label{fig:results_healthy}
\end{figure}

Figure \ref{fig:results_healthy} showcases 3D RI cross-sections of the healthy organoid phantom imaged using the three QPT systems and two RI contrasts. The low-contrast reconstructions (Fig.~\ref{fig:results_healthy}a-c) show the distinct morphology of the organoid wall surrounding the lumen (biologically representing the actin and individual cell nuclei). IDT measurements provided the best signal-to-noise ratio, owing to lower coherent noise than the HT systems. On the other hand, the IDT reconstructed RI values were underestimated when compared to the other tomographs and to the expected value. Both HT reconstructions, while more accurate in terms of RI quantification, suffered from noise derived from the coherent illumination source. Additionally, the NIR reconstruction clearly exhibit lower resolution due to the longer wavelength than the VIS system.
The designed total dry mass was \SI{149}{\nano\gram}, while the measured dry mass was 57\%, 74\% and 152\% of the expected value for the IDT, VIS-HT and NIR-HT systems, respectively. The root mean square error (RMSE) calculated for the plotted RI profile (Fig.~\ref{fig:results_healthy}c) with respect to the designed profile was 0.0026, 0.0017 and 0.0018.

The sample under high RI contrast (Fig.~\ref{fig:results_healthy}d-f) clearly demonstrated reconstruction degradation due to the much higher RI gradients and multiple scattering. The IDT was unable to provide low frequency information leading to underestimated RI values, however, the technique still retrieved the overall geometry and morphology of the sample within the limited numerical aperture. The VIS-HT technique was limited by the high dynamic range of the phase maps, resulting from large optical path differences introduced in the recorded projections of the phantom. This resulted in phase ambiguity leading to unwrapping errors, erroneous projections and subsequently divergent reconstruction. 
Due to this, the morphological features such as the outline of the organoid wall or individual nuclei, were poorly preserved and barely visible. 
NIR-HT was able to mediate this issue thanks to the longer illumination wavelength and effectively provided higher resolution reconstruction than VIS-HT. The RI values were still underestimated by about 75\% globally and 50\% in the analyzed cross-section due to the missing cone and the applicability of the Rytov approximation. The total dry mass calculated from the phantom design was \SI{1223}{\nano\gram}, while the measured dry mass constitute 9\%, 25\% and 51\% of the expected value for the IDT, VIS-HT and NIR-HT systems, respectively. The RMSE in the plotted cross-section (Fig.~\ref{fig:results_healthy}f) was 0.0279, 0.0256 and 0.0150 for the IDT, VIS-HT and NIR-HT systems, respectively.

\begin{figure}[htbp]
    \centering
    \includegraphics[width=1\textwidth]{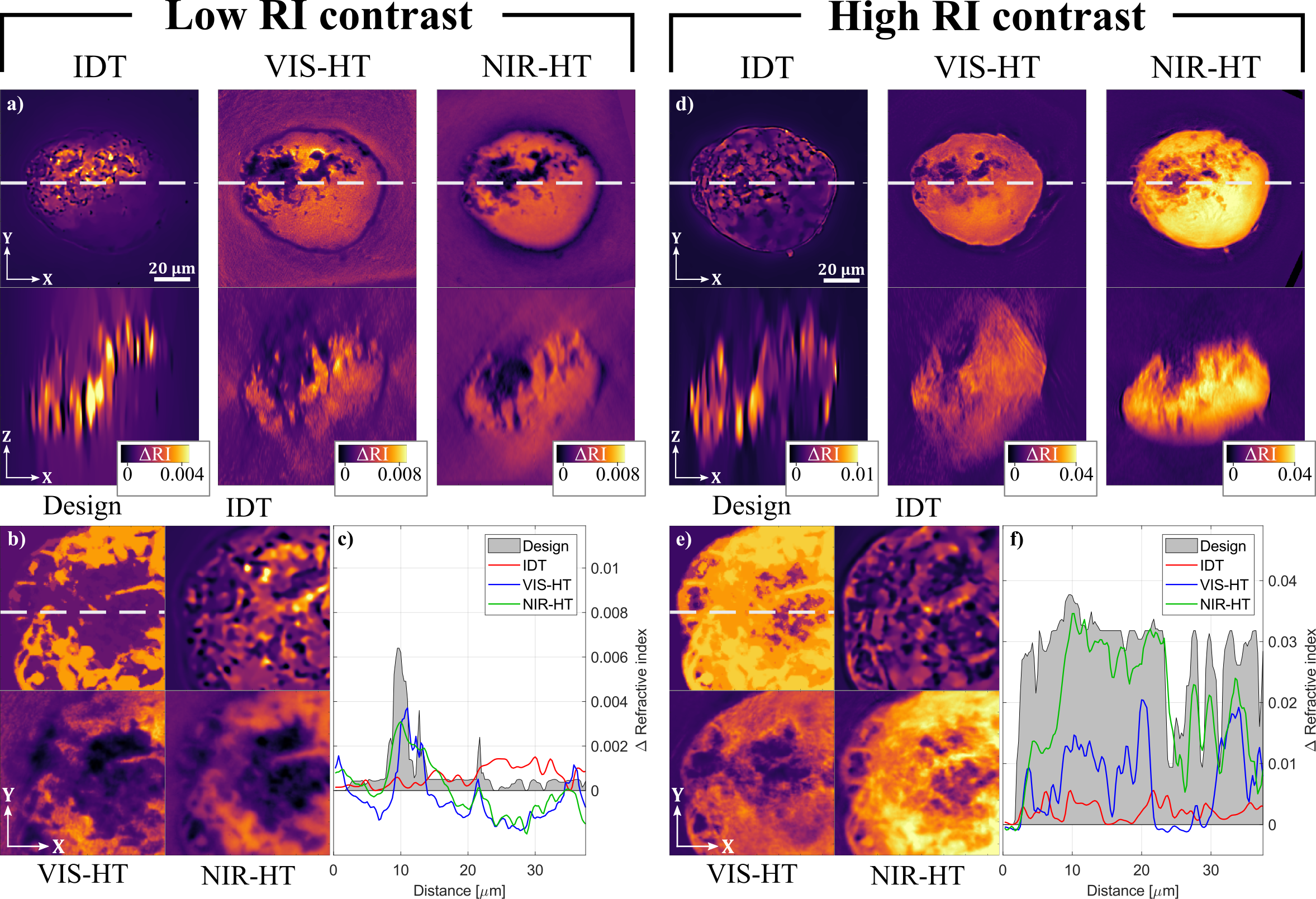} 
    \caption{3D RI reconstructions of a tumoral MLO phantom measured under low (a-c) and high (d-f) RI contrast. Panels a) and d) present lateral and vertical cross-sections for IDT, VIS-HT and NIR-HT systems, while the b) and e) show a region of interest near the central part of the organoid and the RI profile (c, f) across the dashed line in b) and e). Note that the RI colorbar for the IDT system is adjusted for visual contrast and is different from the HTs.}
    \label{fig:results_tumoral}
\end{figure}

The MTLO phantom featured collapsed walls and much more densely packed microstructures (Fig.~\ref{fig:results_tumoral}). In the case of the low contrast sample (Fig.~\ref{fig:results_tumoral}a-c), IDT could not correctly retrieve low-frequency information, thus highlighting regions of high-RI gradients only and lacking major morphological features like overall shape and a cavity shown in the Fig.~\ref{fig:results_tumoral}b. 
The HT reconstructions provided better consistency with the designed RI distribution. Since the tumoral organoid is smaller in the axial (Z) direction than the healthy one and the majority of the RI gradients are concentrated in the center of the organoid, this phantom introduced less disturbances in the passing wavefront and thus is more accurately represented in captured projections, leading to high-fidelity reconstructions. However, the RI plot revealed major discrepancy between the HTs and designed profile, which can be attributed to inaccurate modeling of the flood-polymerization of the low-contrast sample, rather than the reconstruction errors (see Materials and Methods). The total dry mass of the low-RI phantom was \SI{75}{\nano\gram}, while the measured dry mass was equal to 38\%, 98\% and 119\% of the expected value for the IDT, VIS-HT and NIR-HT systems, respectively. Consequently, the RMSE in the plotted cross-section (Fig.~\ref{fig:results_tumoral}c) was 0.0015, 0.0016 and 0.0012.

The high-contrast case (Fig.~\ref{fig:results_tumoral}d-f) highlighted both the weakness of IDT and the advantage of NIR-HT over VIS-HT. The IDT reconstruction underestimated the absolute RI values of the sample, as in the case of the low contrast. Additionally, the morphological details were lost resulting in a noisy  reconstruction without well-defined structures when compared to the original design. VIS-HT was able to retrieve the general features of the organoid, but due to the high dynamic range of the phase maps and subsequent unwrapping errors, the fine details were averaged out and lost. Importantly, NIR-HT was able to overcome these issues and yielded the highest-fidelity reconstruction with the closest match to the designed RI distribution. In the high-contrast MTLO phantom the expected total dry mass was \SI{600}{\nano\gram}, while the measured dry mass constituted 10\%, 60\% and 84\% of the designed value for the IDT, VIS-HT and NIR-HT systems, respectively. The RMSE calculated for the plotted RI profile (Fig.~\ref{fig:results_tumoral}f) with respect to the designed profile was 0.0260, 0.0214 and 0.0105 for the IDT, VIS-HT and NIR-HT, respectively.

\section{Discussion}\label{sec:discussion}
We present here a novel pipeline to create bio-inspired phase-calibrated 3D micro-phantoms for the analysis of healthy liver organoids or those that reflect the pathological state of diet-induced liver cancer. The phantoms were used to characterize different QPT systems and demonstrated their capabilities for metrological analysis tailored for specific biomedical applications. 
Compared to existing phase-calibrated targets, the phantoms described here offer the advantage of mimicking the morphological and functional information found in biological 3D samples. The overall pipeline can be adapted to a wide range of specimens, imaging techniques and modalities in order to create high-fidelity phantoms with features exceeding the expected imaging quality of application-oriented QPT systems. Additionally, RI variations of the sample govern the imaging quality in a vast majority of optical methods, which hints at the value of such phantoms for testing techniques beyond the phase imaging. 

The acquisitions performed using different QPT systems showcased the advantages and drawbacks of the three different phase imaging techniques, highlighting the need for complementary setups to conduct in-depth studies on the optical properties of 3D biological samples. The IDT approach benefits from the compactness of the optical setup at the expenses of RI quantification. The technique could correctly retrieve the overall geometry and morphology of the sample on individual cross-sections, opening up novel directions on organoid time-lapse imaging. Indeed, one of the main challenge in this field is the monitoring of the structural morphology during organoid development. Complementarily, the HT techniques offered more accurate RI quantification capabilities related to the functional and biological integrity of the samples. These data are particularly useful to monitor the dry mass values of the overall organoid and internal structures at individual time-points, giving insights on the evolution of a 3D sample in culture. 

When compared to healthy conditions, tumoral organoid phantoms presented a more challenging scenario in terms of RI and morphology retrieval. These samples featured denser internal reorganization of the actin cytoskeleton, typical of the disease pathology. Under these conditions, the use of a NIR wavelength with reduced light scattering clearly showed a significant improvement in the overall quality of the images as well as RI quantification. Identifying cases where seemingly small morphological changes or measurement parameters can influence the accuracy of the reconstruction, is crucial for assessing the reliability of future experiments and establishing guidelines for QPT users.

The phantoms we developed also allowed the optimization of the reconstruction hyper-parameters and the designing of the data processing chain in advance of an acquisition campaign. Overall, these novel tools gave a deeper understanding of the information content of the RI measurements and provided access to ground-truth values usually unavailable from biological imaging. The design of \emph{ad hoc} phantoms may ultimately help optimizing the hardware setup and reconstruction strategy for a given biological problem. 

\section{Materials and methods}\label{sec:methods}
\textbf{Establishment of mouse models of liver disease.} Mice (strain: C57BL6/J) were housed in isolated light-tight armoires maintained under 12:12hr light:dark cycles and raised on standard chow for 6 weeks. Animals were then switched on to CDAHFD diet (0.1\% methionine, no choline, 60kcal\% fat) for up to 60 weeks. Every 4-8 weeks, animals were sacrificed, livers were isolated and analyzed for liver disease markers by in situ hybridization and histopathology, triglyceride and free fatty-acid levels, fat accumulation, hepatocellular ballooning, inflammation, fibrosis by histology. Organoids were derived from isolated livers as described below.  All animals used in this study were housed at the PEHR, IGFL ENS Lyon. The animal experimentation protocols were submitted and approved by the university ethics committee for animal use and welfare (CECAPP) and approved by the French health ministry (APAFIS\# 2018112114408109 FR \#18197). 

\textbf{Mouse liver organoid culture.} Organoids from mouse liver were cultured following the protocol described in Broutier \emph{et al.} \cite{Broutier2016}. After 7 days of incubation in standard culture conditions, organoids were passaged for maintenance in Hepaticult Organoid Growth Medium (Mouse). Growth medium was changed every 3 to 4 days in culture. For imaging analysis, organoids were either subjected to mechanical disruption or digested into single cells using TrypLE (Gibco Cat: 12605010). Fragments of organoids were resuspended in basement matrix and seeded in 24-well culture plate.

\textbf{Organoid staining and 3D confocal imaging.} Organoid immunofluorescence was performed according to Dekkers \emph{et al.} \cite{Dekkers2019}. 24 hours after seeding, organoids were removed from the Matrigel with cell recovery solution (Corning Cat: 354253). Then, organoids were fixed for 40 minutes into a 4\% PBS-PFA (Thermo Cat:047377) solution and washed for 10 minutes in a PBS-0.1\% Tween20 solution (Sigma Cat: P9416). The primary staining was carried out with EpCAM (CST Cat: 42515), Alpha Smooth Muscle Actin (Novus Cat: NB300-978) and Vimentin (Abcam Cat: Ab20346) diluted 1:500 in PBS-BSA 0,2\% - Triton X100 0,1\% (wash buffer). After an overnight incubation, organoids were washed three times with the wash buffer. The secondary antibodies (Thermo Cat: A21206, A31571 and A11056; 1:2000 dilution) were incubated overnight. Organoids were then serially washed three times in wash buffer before counterstaining the nuclei with Hoechst dye (Cat: 33342). After staining, organoids were retrieved in a clearing solution (Glycerol 60\%, Glucose 2,5M) and mounted with spacers between slide and coverslip. Confocal acquisitions were performed with a Zeiss LSM800 confocal microscope (Laser: \SI{352}{\nano \meter}/\SI{488}{\nano \meter}/\SI{546}{\nano \meter}/\SI{633}{\nano \meter}, Gain: 650 for \SI{352}{\nano \meter} laser and 800 for \SI{488}{\nano \meter}/\SI{546}{\nano \meter}/\SI{633}{\nano \meter} lasers, Z steps: \SI{0.45}{\micro \meter}, Stack thickness: \SI{1}{\micro \meter}).

\textbf{Confocal image processing.} 3D confocal images of both MLO and MTLO were split into individual fluorescent channels (i.e. Hoechst and Actin) and directly segmented in CellPose \cite{stringer_cellpose_2021} using the cyto2 model (the original cell pose model for the cellular cytoplasm segmentation).
To simulate multiple values of RIs within the internal MTLO structure, the actin channel was further thresholded into 3 classes based on different levels of fluorescence intensity.

\textbf{Fabrication and sample preparation for QPT.}
The fabrication was enabled by the two-photon polymerization (TPP) lithography \cite{Wang2023_TPP}. In TPP, a tightly focused femtosecond laser beam induces polymerization of the photosensitive resin near the focal spot by means of two-photon absorption. By scanning the focus across the resin it is possible to additively manufacture the structures line by line, layer by layer with submicrometer resolution in all three dimensions. Most notably, in TPP the degree of monomer cross-linking is proportional to the RI \cite{Zukauskas2015_TPP_GRIN} and depends on the energy dose, therefore the RI distribution within the sample can be engineered by adjusting the process parameters. Fabrication has been performed using Photonic Professional GT2 (Nanoscribe GmbH \& Co. KG, Karlsruhe, Germany) equipped with a 1.4 NA 63x microscope objective, piezo stage for vertical positioning of the sample and dual-axis galvo for lateral scanning. 

Labels segmented from confocal images were assigned the RI values corresponding to the physiological RI distribution and considering the range of values accessible in the resin processed by the TPP. The structures have been sliced and hatched at \SI{0.3} and \SI{0.2}{\micro\meter} respectively in order to define scanning trajectories at a galvo scan speed of \SI{10000}{\micro\meter/\second}. The RI has been modulated locally via the laser power of the TPP system, given as \% of the \SI{50}{\milli\watt} mean optical power entering the aperture of the objective, and calibrated using the method described in \cite{Ziemczonok2023_pw} at \SI{633}{\nano\meter} wavelength. 
The binary mask of the actin channel of the MLO has been dilated (radius of \SI{1.5}{\micro\meter}) in order to properly seal the lumen from the outside of the organoid and increase structural stability, as it is essentially a sphere with a relatively thin wall that could collapse during development. The Hoechst and Actin 3D masks were assigned the RI of 1.5348 and 1.5497 respectively, which correspond to the laser power of 20 and 40\%. In case of MTLO, the phantom has been assigned the RI of 1.5348, 1.5443 and 1.5497 for the three Actin masks of increasing fluorescence intensity, which correspond to the fabrication laser powers of 20, 25 and 40\%.

The phantoms have been fabricated out of the IP-Dip2 resin (Nanoscribe) on top of a \#1.5H coverslip. After fabrication, the sample intended for high RI contrast has been developed in PGMEA (Propylene glycol monomethyl ether acetate; 12 min; Sigma Cat: 484431) followed by isopropyl alcohol (10 min; Sigma Cat: 190764) and then blow-dried. Then, it was immersed in a drop of immersion medium providing desired RI contrast, in this case 0.0374 (Immersol 518F, Carl Zeiss AG, Oberkochen, Germany), and covered with a \qtyproduct[product-units = single]{24 x 60 x 0.17}{\milli\meter} coverslip with a \SI{120}{\micro\meter} thick imaging spacer in between (SS1X9, Grace Bio-Labs, Bend, Oregon, USA). The low RI contrast sample has not been developed at all, instead the surrounding unpolymerized resin has been flood-polymerized using white-light LED. In this case the maximal RI contrast between the TPP-polymerized structure and the partially-polymerized surrounding resin has been measured using the calibration step structure printed alongside the organoids and was found to be 0.0061.

\textbf{QPT results processing and comparison.} 
The QPT reconstructions have been resampled to \qtyproduct[product-units = single]{0.25 x 0.25 x 1}{\micro\meter} per voxel (linear interpolation), translated, rotated (linear interpolation), and then cropped in order to unify their size, sampling, position and orientation, as well as to align their features (manual registration). 
The 3D RI design has been obtained from the given fabrication design with RI values corresponding to the local laser powers (see Supplementary for details). The outermost region has been flood-filled with the immersion medium RI. 
The expected 3D RI does not account for fabrication inaccuracies such as shrinkage, as it is highly anisotropic, or excessive development of the partially-polymerized regions being in contact with the developer. Additionally, in the case of low-contrast 3D RI design there were major difference between the expected and measured RI profiles. This is likely due to the liquid resin around and inside the microstructures being photobleached, and therefore flood-polymerization was less effective and resulted in lower RI in experimental samples. 
These are the pain-points of the presented methodology which should be considered or minimized during the design stage. 
The dry mass equivalent contained in the whole reconstruction volume expressed in nanograms is equal to the sum of $\Delta RI \Delta x \Delta y \Delta z/(\alpha \cdot 1000)$ over all voxels, where $\Delta x$, $\Delta y$ and $\Delta z$ denote voxel size and $\alpha$ is a refractive index increment equal to \SI{0.18}{\micro\meter^3\per\pico\gram}. The RMSE is calculated as a square root of the mean of the $RI_{measured} - RI_{designed}$ values along the analyzed RI profile in registered reconstructions. 

\textbf{Intensity diffraction tomography.} 
IDT images were collected as previously described \cite{Pierre:22}. Briefly, a commercial LED array (Sci-Microscopy – $\lambda =$ 532 nm) was used to subsequently illuminate the samples at different angles with a 40$\times$ Olympus objective 0.65 NA. Transmitted photons are then focused on a CMOS sensor (IDS, UI-3880CP-M-GL) using a tube lens of 100 mm (Thorlabs, TTl100-A). 85 images were acquired with exposure times calibrated for each angle of the LED array (acquisition time: in the range of 80-800 ms, maximum illumination angle: 10°). 
The 3D objects were reconstructed using custom scripts developed in Python with the methodology described in Pierr\'e {\it et al.} \cite{Pierre:22}. Briefly, the algorithm consists of a multi-slice scattering model (beam propagation method \cite{VanRoey:81}) coded inside a deep-learning framework (diffractive deep neural network, DDNN). The optimized weights of the DDNN are the RIs and absorption of the sample, calculated layer by layer on the sliced object. Three different network layers were used to model the light diffraction: an FFT, a product, and an exponential product-layer. The propagation of the field through the object alternates between the Fourier and the signal space. After the 2D field propagation slice by slice, the reconstruction of the image is obtained with the training of the DDNN with total variation regularization. The optimization is performed with a fast version of the iterative shrinkage-thresholding algorithms (FISTA) \cite{kamilov_optical_2016}. Error backpropagation is achieved with a PyTorch automatic differentiation feature. Reconstructions (40 iterations) last approximately 15 min for a 1024$\times$1024$\times$80 volume.
The regularization parameters play crucial part in the reconstruction fidelity. We used MLO phantom at low RI contrast to evaluate the parameters of the optimization criterion during the 3D reconstruction for the IDT system (see Supplementary). 

\textbf{VIS-HT}
The visible-light HT system uses a 633 nm single longitudinal mode volume-Bragg-grating-stabilized laser source and is based on a Mach-Zehnder interferometer equipped with a typical illumination and imaging module at numerical aperture NA=1.3. Transverse magnification in the imaging part is 48.5 and the camera pixel size is $3.45 \mu m$. In this work the system works with 90 projections at annular illumination scenario. The sample illumination angle is up to $45^{o}$. After capturing data, the complex amplitudes are retrieved with the Fourier transform method, and the 3D tomographic reconstructions of the RI distribution are calculated with the open-source EWALD software (see Data Availability section) that uses Fourier diffraction theorem \cite{kus2019holographic} with the Rytov approximation. Specifically, Direct Inversion method is used where each Rytov-approximated projection is Fourier transformed and is cast onto the Ewald sphere in the empty matrix representing spectrum of the tomographic reconstruction. The radius of the Ewald sphere depends on the illumination wavelength and the RI of the immersion, whereas its position depends on the illumination angle. The values of pixels in the Fourier spectrum that are filled multiple times by projections are averaged to avoid low frequencies being overrepresented. After all projections are processed, the inverse Fourier transform is calculated from the filled spectrum to obtain scattering potential of the sample, from which the 3D RI distribution can be directly calculated.

\textbf{NIR-HT}
The NIR-HT system is a Mach-Zehnder-based tomographic microscope, modified to incorporate an optical path equalizer. This is a part required to provide high contrast interference pattern for a source with shorter temporal coherence than single longitudinal mode lasers used in VIS-HT. The illumination was provided by the swept-source laser, operating in a single wavelength mode at \SI{878}{\nano \meter}, with the full width at half maximum (FWHM) of \SI{0.06}{\nano \meter}. This setup utilizes 40x NA=1.3 microscope objectives and the effective magnification of the imaging part is 72.7. The pixel of the camera used in the setup was $5.5\mu m$. In total, 181 projections are captured during the measurement process - 180 illuminations around a circle in the pupil plus one axial illumination direction. After capturing data, projections with erroneous phase information were removed from the sinogram. Finally, the tomographic reconstructions were calculated with the same algorithm as in the case of VIS-HT.

\backmatter
\bmhead{Data availability statement}
Designed and measured 3D refractive index volumes, as well as corresponding raw measurement files are available at \url{https://doi.org/10.5281/zenodo.11242045}. HT reconstruction algorithm is open-source and available at \url{https://github.com/biopto/EWALD}.

\bmhead{Acknowledgments}
This work has received funding from the European Union’s Horizon 2020 research program under grant agreement N°101016726. This work has also been funded in part by the French National Research Agency (ANR), project LIVE 3D-CNN (ANR-21-CE19-0020), Polish Ministry of Education and Science (Polish Metrology, PM/SP/0079/2021/1) and Warsaw University of Technology under the program Excellence Initiative: Research University (IDUB). The authors would like to thank Dr. C. Acquitter for the discussions over 3D segmentation. MZ acknowledges the support by the Foundation for Polish Science (FNP).

\bmhead{Author Contribution}
\textbf{MZ:} Conceptualization, Methodology, Data curation, Formal analysis, Visualization, Writing – original draft. 
\textbf{SD:} Methodology, Data curation, Visualization, Writing – review \& editing.
\textbf{JN:} Data curation, Visualization, Writing – original draft.
\textbf{AK:} Conceptualization, Investigation, Data curation, Writing – review \& editing.
\textbf{LH:} Software, Formal analysis, Writing – review \& editing.
\textbf{CF:} Software.
\textbf{GG:} Software.
\textbf{MF:} Resources.
\textbf{DS:} Resources.
\textbf{KP:} Supervision, Writing – review \& editing, Funding acquisition.
\textbf{CP:} Conceptualization, Methodology, Visualization, Writing – original draft, Project administration, Funding acquisition.
\textbf{WK:} Writing – review \& editing, Project administration, Funding acquisition.
\textbf{MK:} Writing – review \& editing, Project administration, Funding acquisition.

\bmhead{Conflict of interest}
MZ, AK and MK are named inventors on the patent application which is tangential to the subject of this work (PCT/IB2020/054772). Remaining authors declare no conflicts of interest.

\bibliography{sn-bibliography}

\begin{thebibliography}{10}
\expandafter\ifx\csname url\endcsname\relax
  \def\url#1{\burl{#1}}\fi
\expandafter\ifx\csname urlprefix\endcsname\relax\def\urlprefix{URL }\fi
\providecommand{\bibinfo}[2]{#2}
\providecommand{\eprint}[2][]{\url{#2}}
\providecommand{\doi}[1]{\url{https://doi.org/#1}}
\bibcommenthead

\bibitem{zhao_organoids_2022}
\bibinfo{author}{Zhao, Z.} \emph{et~al.}
\newblock \bibinfo{title}{Organoids}.
\newblock \emph{\bibinfo{journal}{Nature Reviews Methods Primers}} \textbf{\bibinfo{volume}{2}}, \bibinfo{pages}{94} (\bibinfo{year}{2022}).

\bibitem{Duval2017_2Dvs3D_cell_culture}
\bibinfo{author}{Duval, K.} \emph{et~al.}
\newblock \bibinfo{title}{Modeling physiological events in {2D} vs. {3D} cell culture}.
\newblock \emph{\bibinfo{journal}{Physiology}} \textbf{\bibinfo{volume}{32}}, \bibinfo{pages}{266--277} (\bibinfo{year}{2017}).

\bibitem{susaki_perspective_2021}
\bibinfo{author}{Susaki, E.~A.} \& \bibinfo{author}{Takasato, M.}
\newblock \bibinfo{title}{Perspective: extending the utility of three-dimensional organoids by tissue clearing technologies}.
\newblock \emph{\bibinfo{journal}{Frontiers in Cell and Developmental Biology}} \textbf{\bibinfo{volume}{9}}, \bibinfo{pages}{679226} (\bibinfo{year}{2021}).

\bibitem{Pawley2005}
\bibinfo{author}{Pawley, J.}
\newblock \emph{\bibinfo{title}{Handbook of biological confocal microscopy}} Vol. \bibinfo{volume}{236} (\bibinfo{publisher}{Springer Science \& Business Media}, \bibinfo{year}{2006}).

\bibitem{Astratov2023_roadmap_on_labelfree}
\bibinfo{author}{Astratov, V.~N.} \emph{et~al.}
\newblock \bibinfo{title}{Roadmap on label-free super-resolution imaging}.
\newblock \emph{\bibinfo{journal}{Laser \& Photonics Reviews}} \textbf{\bibinfo{volume}{17}}, \bibinfo{pages}{2200029} (\bibinfo{year}{2023}).

\bibitem{nguyen_quantitative_2022}
\bibinfo{author}{Nguyen, T.~L.} \emph{et~al.}
\newblock \bibinfo{title}{Quantitative phase imaging: recent advances and expanding potential in biomedicine}.
\newblock \emph{\bibinfo{journal}{ACS Nano}} \textbf{\bibinfo{volume}{16}}, \bibinfo{pages}{11516--11544} (\bibinfo{year}{2022}).

\bibitem{park_quantitative_2018}
\bibinfo{author}{Park, Y.}, \bibinfo{author}{Depeursinge, C.} \& \bibinfo{author}{Popescu, G.}
\newblock \bibinfo{title}{Quantitative phase imaging in biomedicine}.
\newblock \emph{\bibinfo{journal}{Nature Photonics}} \textbf{\bibinfo{volume}{12}}, \bibinfo{pages}{578--589} (\bibinfo{year}{2018}).

\bibitem{Verrier2024_ODT_review}
\bibinfo{author}{Verrier, N.}, \bibinfo{author}{Debailleul, M.} \& \bibinfo{author}{Haeberl{\'e}, O.}
\newblock \bibinfo{title}{Recent advances and current trends in transmission tomographic diffraction microscopy}.
\newblock \emph{\bibinfo{journal}{Sensors}} \textbf{\bibinfo{volume}{24}}, \bibinfo{pages}{1594} (\bibinfo{year}{2024}).

\bibitem{lim2019high}
\bibinfo{author}{Lim, J.}, \bibinfo{author}{Ayoub, A.~B.}, \bibinfo{author}{Antoine, E.~E.} \& \bibinfo{author}{Psaltis, D.}
\newblock \bibinfo{title}{High-fidelity optical diffraction tomography of multiple scattering samples}.
\newblock \emph{\bibinfo{journal}{Light: Science \& Applications}} \textbf{\bibinfo{volume}{8}}, \bibinfo{pages}{82} (\bibinfo{year}{2019}).

\bibitem{chen2020multi}
\bibinfo{author}{Chen, M.}, \bibinfo{author}{Ren, D.}, \bibinfo{author}{Liu, H.-Y.}, \bibinfo{author}{Chowdhury, S.} \& \bibinfo{author}{Waller, L.}
\newblock \bibinfo{title}{Multi-layer born multiple-scattering model for 3d phase microscopy}.
\newblock \emph{\bibinfo{journal}{Optica}} \textbf{\bibinfo{volume}{7}}, \bibinfo{pages}{394--403} (\bibinfo{year}{2020}).

\bibitem{pham2020three}
\bibinfo{author}{Pham, T.-a.} \emph{et~al.}
\newblock \bibinfo{title}{Three-dimensional optical diffraction tomography with lippmann-schwinger model}.
\newblock \emph{\bibinfo{journal}{IEEE Transactions on Computational Imaging}} \textbf{\bibinfo{volume}{6}}, \bibinfo{pages}{727--738} (\bibinfo{year}{2020}).

\bibitem{Ossowski2022}
\bibinfo{author}{Ossowski, P.} \emph{et~al.}
\newblock \bibinfo{title}{Near-infrared, wavelength, and illumination scanning holographic tomography}.
\newblock \emph{\bibinfo{journal}{Biomedical Optics Express}} \textbf{\bibinfo{volume}{13}}, \bibinfo{pages}{5971--5988} (\bibinfo{year}{2022}).

\bibitem{Lee2023}
\bibinfo{author}{Lee, M.~J.} \emph{et~al.}
\newblock \bibinfo{title}{Long-term three-dimensional high-resolution imaging of live unlabeled small intestinal organoids using low-coherence holotomography}.
\newblock \emph{\bibinfo{journal}{bioRxiv}} \bibinfo{pages}{2023--09} (\bibinfo{year}{2023}).

\bibitem{chowdhury_high-resolution_2019}
\bibinfo{author}{Chowdhury, S.} \emph{et~al.}
\newblock \bibinfo{title}{High-resolution {3D} refractive index microscopy of multiple-scattering samples from intensity images}.
\newblock \emph{\bibinfo{journal}{Optica}} \textbf{\bibinfo{volume}{6}}, \bibinfo{pages}{1211--1219} (\bibinfo{year}{2019}).

\bibitem{tian_3d_2015}
\bibinfo{author}{Tian, L.} \& \bibinfo{author}{Waller, L.}
\newblock \bibinfo{title}{{3D} intensity and phase imaging from light field measurements in an {LED} array microscope}.
\newblock \emph{\bibinfo{journal}{Optica}} \textbf{\bibinfo{volume}{2}}, \bibinfo{pages}{104--111} (\bibinfo{year}{2015}).

\bibitem{krauze20223d}
\bibinfo{author}{Krauze, W.} \emph{et~al.}
\newblock \bibinfo{title}{{3D} scattering microphantom sample to assess quantitative accuracy in tomographic phase microscopy techniques}.
\newblock \emph{\bibinfo{journal}{Scientific Reports}} \textbf{\bibinfo{volume}{12}}, \bibinfo{pages}{19586} (\bibinfo{year}{2022}).

\bibitem{Ikawa-Yoshida2017}
\bibinfo{author}{Ikawa-Yoshida, A.} \emph{et~al.}
\newblock \bibinfo{title}{Hepatocellular carcinoma in a mouse model fed a choline-deficient, l-amino acid-defined, high-fat diet}.
\newblock \emph{\bibinfo{journal}{International Journal of Experimental Pathology}} \textbf{\bibinfo{volume}{98}}, \bibinfo{pages}{221--233} (\bibinfo{year}{2017}).
\newblock \urlprefix\url{https://onlinelibrary.wiley.com/doi/abs/10.1111/iep.12240}.

\bibitem{Tang2020}
\bibinfo{author}{Tang, S.} \emph{et~al.}
\newblock \bibinfo{title}{Vital and distinct roles of h2a.z isoforms in hepatocellular carcinoma}.
\newblock \emph{\bibinfo{journal}{OncoTargets and Therapy}} \textbf{\bibinfo{volume}{Volume 13}}, \bibinfo{pages}{4319–4337} (\bibinfo{year}{2020}).
\newblock \urlprefix\url{http://dx.doi.org/10.2147/OTT.S243823}.

\bibitem{lalone_quantitative_2023}
\bibinfo{author}{LaLone, V.} \emph{et~al.}
\newblock \bibinfo{title}{Quantitative chemometric phenotyping of three-dimensional liver organoids by {Raman} spectral imaging}.
\newblock \emph{\bibinfo{journal}{Cell Rep Methods}} \textbf{\bibinfo{volume}{3}}, \bibinfo{pages}{100440} (\bibinfo{year}{2023}).

\bibitem{Liu2016_cellRI}
\bibinfo{author}{Liu, P.~Y.} \emph{et~al.}
\newblock \bibinfo{title}{Cell refractive index for cell biology and disease diagnosis: past, present and future}.
\newblock \emph{\bibinfo{journal}{Lab on a Chip}} \textbf{\bibinfo{volume}{16}}, \bibinfo{pages}{634--644} (\bibinfo{year}{2016}).

\bibitem{Gul2021_cellRI}
\bibinfo{author}{Gul, B.}, \bibinfo{author}{Ashraf, S.}, \bibinfo{author}{Khan, S.}, \bibinfo{author}{Nisar, H.} \& \bibinfo{author}{Ahmad, I.}
\newblock \bibinfo{title}{Cell refractive index: Models, insights, applications and future perspectives}.
\newblock \emph{\bibinfo{journal}{Photodiagnosis and Photodynamic Therapy}} \textbf{\bibinfo{volume}{33}}, \bibinfo{pages}{102096} (\bibinfo{year}{2021}).

\bibitem{Nguyen2022_cellRI}
\bibinfo{author}{Nguyen, T.~L.} \emph{et~al.}
\newblock \bibinfo{title}{Quantitative phase imaging: recent advances and expanding potential in biomedicine}.
\newblock \emph{\bibinfo{journal}{ACS Nano}} \textbf{\bibinfo{volume}{16}}, \bibinfo{pages}{11516--11544} (\bibinfo{year}{2022}).

\bibitem{Yoon2020_scattering_review}
\bibinfo{author}{Yoon, S.} \emph{et~al.}
\newblock \bibinfo{title}{Deep optical imaging within complex scattering media}.
\newblock \emph{\bibinfo{journal}{Nature Reviews Physics}} \textbf{\bibinfo{volume}{2}}, \bibinfo{pages}{141--158} (\bibinfo{year}{2020}).

\bibitem{Lambrou2021_scattering_review}
\bibinfo{author}{Lambrou, G.~I.}, \bibinfo{author}{Tagka, A.}, \bibinfo{author}{Kotoulas, A.}, \bibinfo{author}{Chatziioannou, A.} \& \bibinfo{author}{Matsopoulos, G.~K.}
\newblock \bibinfo{title}{Physical and methodological perspectives on the optical properties of biological samples: {A} review}.
\newblock \emph{\bibinfo{journal}{Photonics}} \textbf{\bibinfo{volume}{8}}, \bibinfo{pages}{540} (\bibinfo{year}{2021}).

\bibitem{Boothe2017_RImatching_medium}
\bibinfo{author}{Boothe, T.} \emph{et~al.}
\newblock \bibinfo{title}{A tunable refractive index matching medium for live imaging cells, tissues and model organisms}.
\newblock \emph{\bibinfo{journal}{Elife}} \textbf{\bibinfo{volume}{6}}, \bibinfo{pages}{e27240} (\bibinfo{year}{2017}).

\bibitem{Lee2022_immersionRImatching}
\bibinfo{author}{Lee, D.} \emph{et~al.}
\newblock \bibinfo{title}{High-fidelity optical diffraction tomography of live organisms using iodixanol refractive index matching}.
\newblock \emph{\bibinfo{journal}{Biomedical Optics Express}} \textbf{\bibinfo{volume}{13}}, \bibinfo{pages}{6404--6415} (\bibinfo{year}{2022}).

\bibitem{Pierre:22}
\bibinfo{author}{Pierr{\'e}, W.} \emph{et~al.}
\newblock \bibinfo{title}{{3D} time-lapse imaging of a mouse embryo using intensity diffraction tomography embedded inside a deep learning framework}.
\newblock \emph{\bibinfo{journal}{Applied Optics}} \textbf{\bibinfo{volume}{61}}, \bibinfo{pages}{3337--3348} (\bibinfo{year}{2022}).

\bibitem{Broutier2016}
\bibinfo{author}{Broutier, L.} \emph{et~al.}
\newblock \bibinfo{title}{Culture and establishment of self-renewing human and mouse adult liver and pancreas {3D} organoids and their genetic manipulation}.
\newblock \emph{\bibinfo{journal}{Nat. Protoc.}} \textbf{\bibinfo{volume}{11}}, \bibinfo{pages}{1724--1743} (\bibinfo{year}{2016}).

\bibitem{Dekkers2019}
\bibinfo{author}{Dekkers, J.~F.} \emph{et~al.}
\newblock \bibinfo{title}{High-resolution 3d imaging of fixed and cleared organoids}.
\newblock \emph{\bibinfo{journal}{Nature Protocols}} \textbf{\bibinfo{volume}{14}}, \bibinfo{pages}{1756–1771} (\bibinfo{year}{2019}).
\newblock \urlprefix\url{http://dx.doi.org/10.1038/s41596-019-0160-8}.

\bibitem{stringer_cellpose_2021}
\bibinfo{author}{Stringer, C.}, \bibinfo{author}{Wang, T.}, \bibinfo{author}{Michaelos, M.} \& \bibinfo{author}{Pachitariu, M.}
\newblock \bibinfo{title}{Cellpose: a generalist algorithm for cellular segmentation}.
\newblock \emph{\bibinfo{journal}{Nature Methods}} \textbf{\bibinfo{volume}{18}}, \bibinfo{pages}{100--106} (\bibinfo{year}{2021}).

\bibitem{Wang2023_TPP}
\bibinfo{author}{Wang, H.} \emph{et~al.}
\newblock \bibinfo{title}{Two-photon polymerization lithography for optics and photonics: fundamentals, materials, technologies, and applications}.
\newblock \emph{\bibinfo{journal}{Advanced Functional Materials}} \textbf{\bibinfo{volume}{33}}, \bibinfo{pages}{2214211} (\bibinfo{year}{2023}).

\bibitem{Zukauskas2015_TPP_GRIN}
\bibinfo{author}{{\v{Z}}ukauskas, A.} \emph{et~al.}
\newblock \bibinfo{title}{Tuning the refractive index in {3D} direct laser writing lithography: towards {GRIN} microoptics}.
\newblock \emph{\bibinfo{journal}{Laser \& Photonics Reviews}} \textbf{\bibinfo{volume}{9}}, \bibinfo{pages}{706--712} (\bibinfo{year}{2015}).

\bibitem{Ziemczonok2023_pw}
\bibinfo{author}{Ziemczonok, M.} \& \bibinfo{author}{Kujawi{\'n}ska, M.}
\newblock \bibinfo{title}{Multiscale and multipurpose phantoms for {2D/3D} quantitative phase imaging}.
\newblock \emph{\bibinfo{journal}{Proc. SPIE}} \textbf{\bibinfo{volume}{1238908}}, \bibinfo{pages}{36--42} (\bibinfo{year}{2023}).

\bibitem{VanRoey:81}
\bibinfo{author}{Van~Roey, J.}, \bibinfo{author}{Van~der Donk, J.} \& \bibinfo{author}{Lagasse, P.}
\newblock \bibinfo{title}{Beam-propagation method: analysis and assessment}.
\newblock \emph{\bibinfo{journal}{Journal of the Optical Society of America}} \textbf{\bibinfo{volume}{71}}, \bibinfo{pages}{803--810} (\bibinfo{year}{1981}).

\bibitem{kamilov_optical_2016}
\bibinfo{author}{Kamilov, U.~S.} \emph{et~al.}
\newblock \bibinfo{title}{Optical tomographic image reconstruction based on beam propagation and sparse regularization}.
\newblock \emph{\bibinfo{journal}{IEEE Transactions on Computational Imaging}} \textbf{\bibinfo{volume}{2}}, \bibinfo{pages}{59--70} (\bibinfo{year}{2016}).

\bibitem{kus2019holographic}
\bibinfo{author}{Ku{\'s}, A.}, \bibinfo{author}{Krauze, W.}, \bibinfo{author}{Makowski, P.~L.} \& \bibinfo{author}{Kujawi{\'n}ska, M.}
\newblock \bibinfo{title}{Holographic tomography: hardware and software solutions for 3d quantitative biomedical imaging}.
\newblock \emph{\bibinfo{journal}{Etri Journal}} \textbf{\bibinfo{volume}{41}}, \bibinfo{pages}{61--72} (\bibinfo{year}{2019}).

\bibitem{Gissibl2017_ipdipRI}
\bibinfo{author}{Gissibl, T.}, \bibinfo{author}{Wagner, S.}, \bibinfo{author}{Sykora, J.}, \bibinfo{author}{Schmid, M.} \& \bibinfo{author}{Giessen, H.}
\newblock \bibinfo{title}{Refractive index measurements of photo-resists for three-dimensional direct laser writing}.
\newblock \emph{\bibinfo{journal}{Optical Materials Express}} \textbf{\bibinfo{volume}{7}}, \bibinfo{pages}{2293--2298} (\bibinfo{year}{2017}).

\end{thebibliography}

\newpage
\begin{appendices}
\makebox[\textwidth]{\textbf{Supplementary}}

\textbf{Immunofluorescence confocal images}
\begin{figure}[htbp]
    \centering
    \includegraphics[width=.7\textwidth]{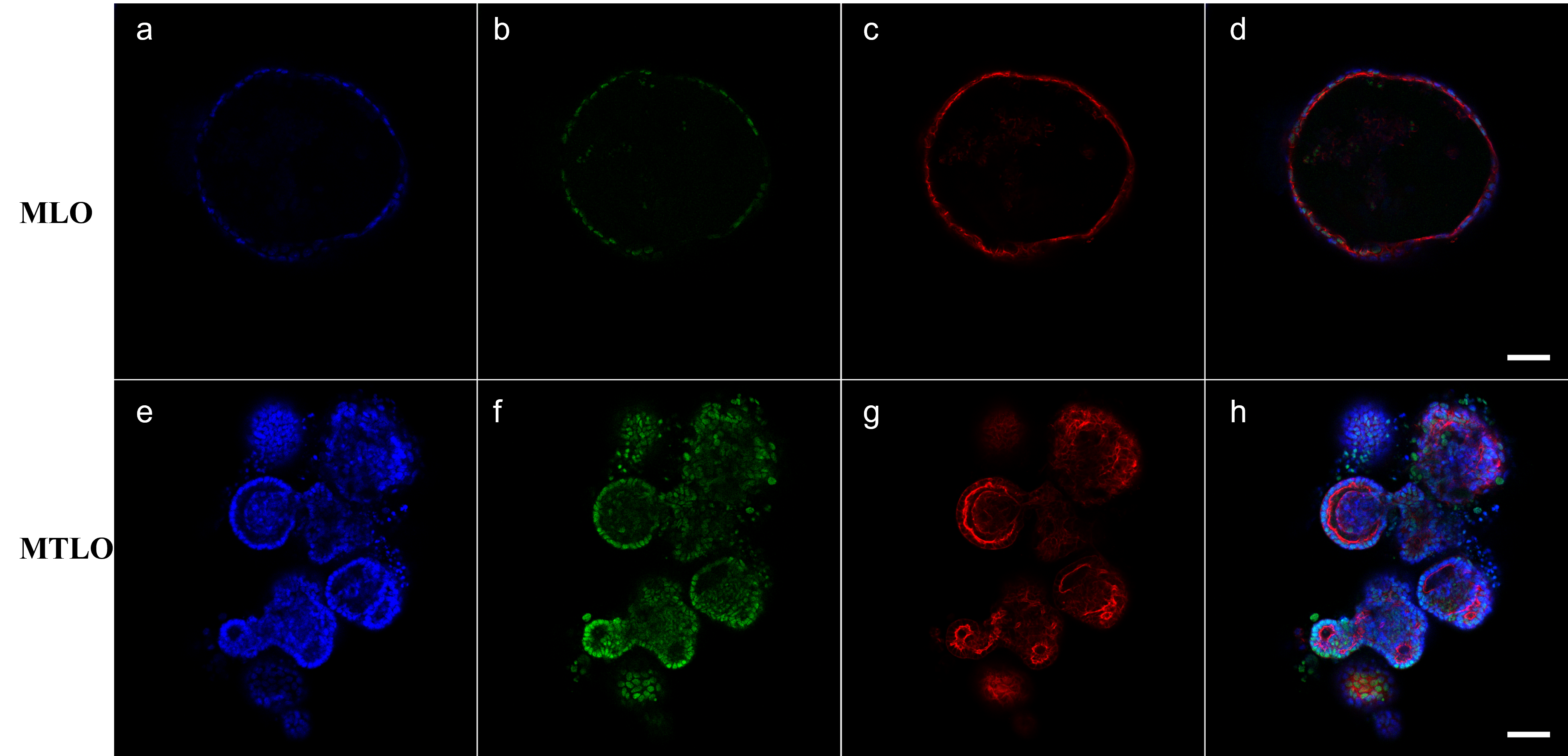} 
    \caption{Immunofluorescence characterization of MLO (a-d) and MTLO (e-f). Fixed organoids were stained with anti H2AZ antibody (b,f), phalloïdin (c,g) and counterstained with Hoechst DNA dye (a,e). Panels d, h are merged images of all three channels respectively. Scale bar is 50µm.}
    \label{fig:sup_fluo}
\end{figure}

\textbf{Tomographic solver input data}
\begin{figure}[h]
    \centering
    \includegraphics[width=1\textwidth]{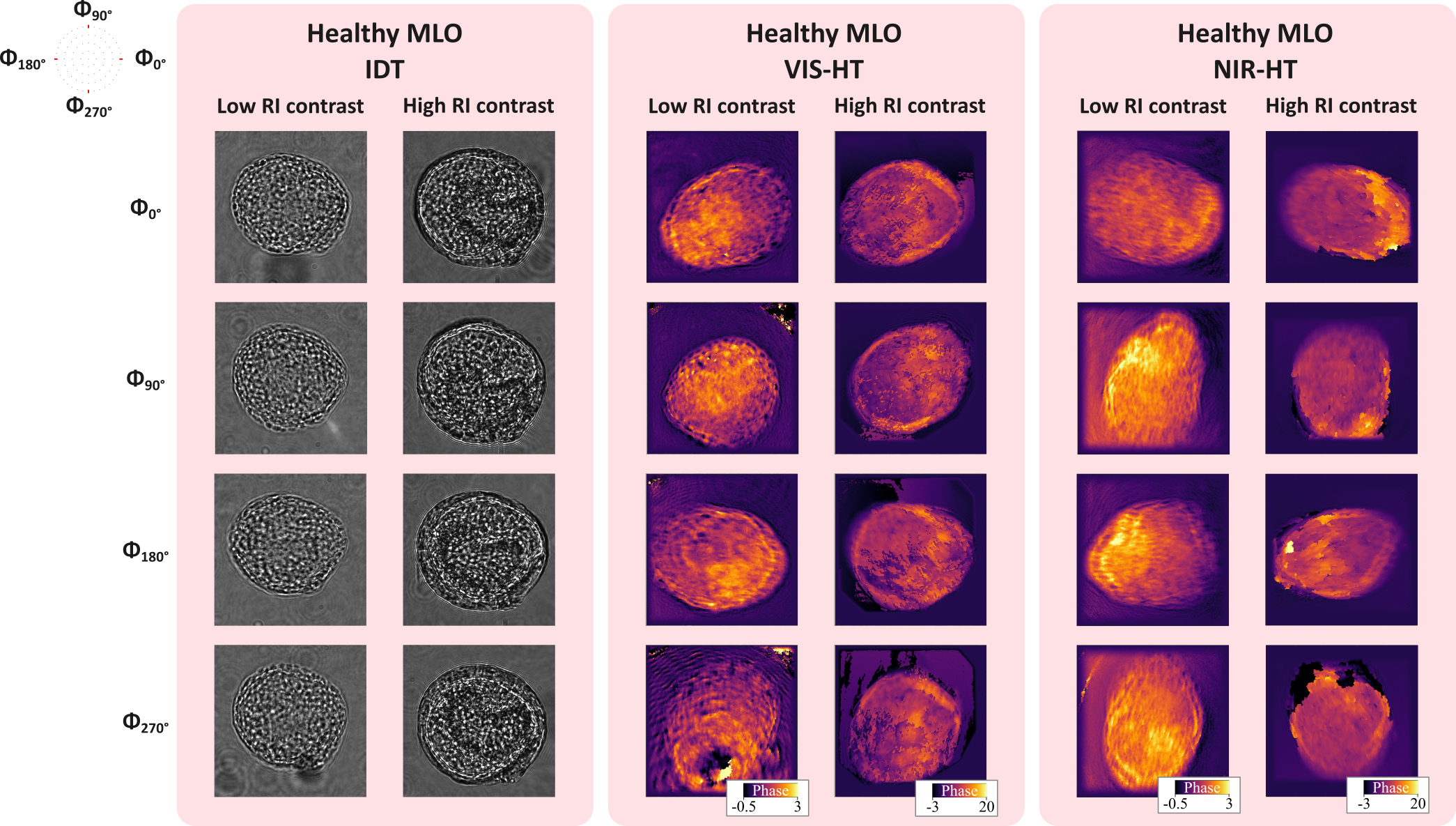} 
    \caption{MLO projections used for the tomographic reconstruction: raw intensity images for IDT and phase maps for HT, captured at different illumination angles and RI contrasts.}
    \label{fig:sup_rawMLO}
\end{figure}

\begin{figure}[h]
    \centering
    \includegraphics[width=1\textwidth]{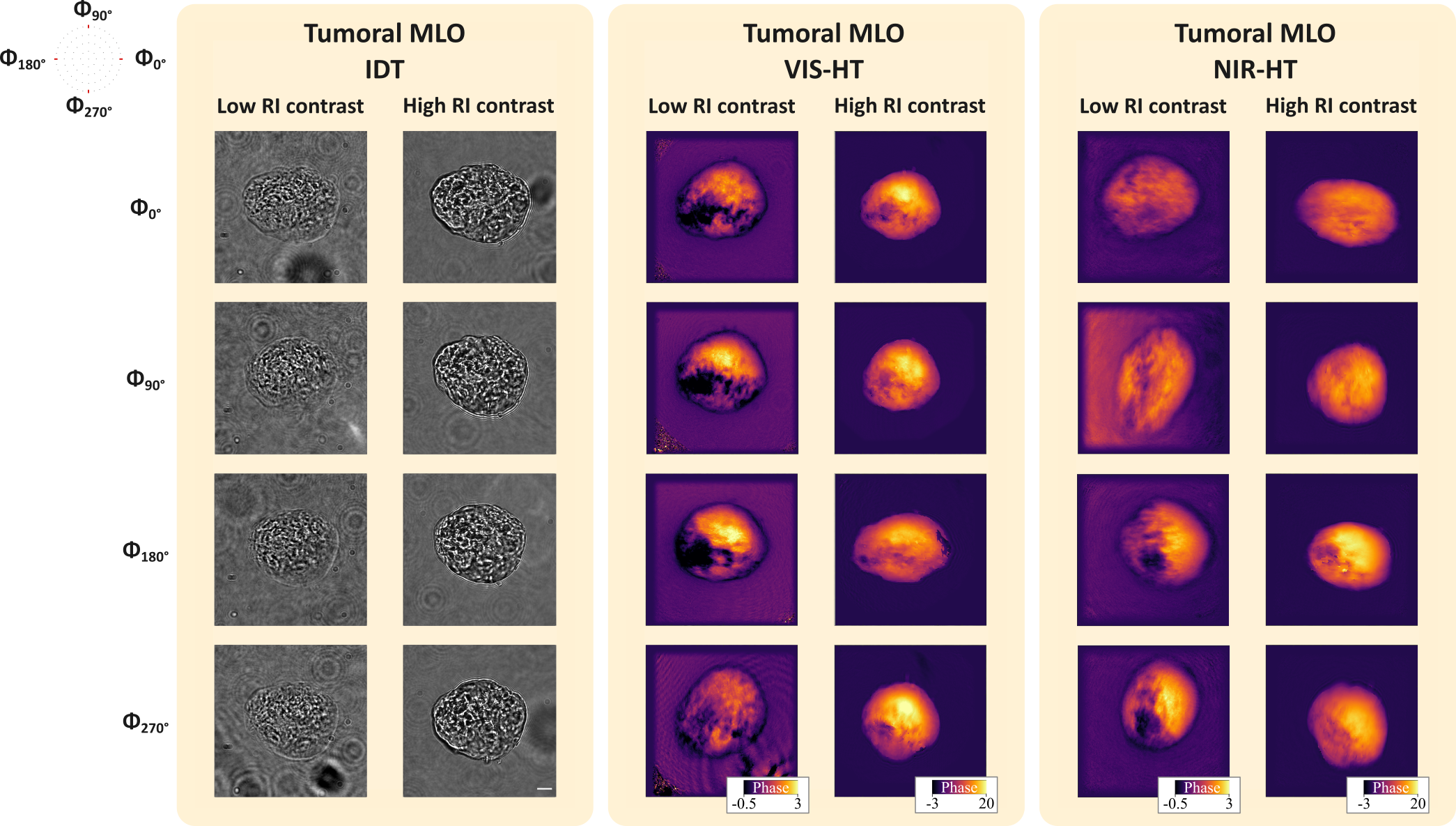} 
    \caption{MTLO projections used for the tomographic reconstruction: raw intensity images for IDT and phase maps for HT, captured at different illumination angles and RI contrasts.}
    \label{fig:sup_rawMTLO}
\end{figure}
\clearpage
\textbf{IDT optimization}
IDT data-processing consists in finding the optical properties of a sample from a set (indexed by $j\le J$) of intensity-only diffraction measurements. Here, the sample was simply illuminated by an incident light, assumed as a monochromatic plane wave with wavelength $\lambda$ and various illumination wave vectors $\vec \mu_j$.

The problem was posed as an inverse problem consisting in: \emph{i}) formulating the physics underlying the experiment, \emph{ii}) formulating a criterion for minimization, such criterion being composed by a data fidelity term and a regularization term, and \emph{iii}) performing the optimization. 

About the physical description of the experiment, light is described by a scalar complex field $A(\vec{r},z)$ where the sample is represented by the real part $\mathcal{R}(\vec{r},z)$ (or $\Delta RI$) and the imaginary part $\mathcal{I}(\vec{r},z)$ of its complex optical index difference to the background optical index $n_0$> The position in space is $\vec{r}$ for \emph{xy}-axes and $z$ for \emph{z}-axis, \emph{z} being chosen as the principal axis of propagation of light. The BPM is used as the model that describes the light propagation inside a medium with arbitrary optical index; it consists in diving the medium in $P$ slices, with slice to slice distance $\Delta z$. Finally, noting $A^-_p$ and $A^+_p$ the optical fields before and after slice $p$, the BPM model formulation is:
\[
\left\{
\begin{array}{lc}
A^-_1(\vec{r}) = exp(2i\pi\vec{\mu}_j.\vec{r})&(\text{illumination})\\
\forall p\le P, A^+_p = A^-_p.exp \left( \frac{2i\pi\Delta z}{\lambda}\left(\mathcal{R}_p+i\mathcal{I}_p\right) \right)&(\text{transmission})\\
\forall p<P, A^-_{p+1}=FT^{-1}\left[FT\left[A^+_{p}\right].H_{\Delta z}\right]&(\text{propagation})\\
\end{array}
\right.
\]
where $FT\left[.\right]$ and $FT^{-1}\left[.\right]$ are the Fourier transform and the inverse Fourier transform and where $H_z(\vec\mu)=exp(2i\pi z\sqrt{n_0^2/\lambda^2-\vec\mu^2})$ is the ``angular-spectrum" kernel which allows to propagate the optical field by a distance $z$ in a medium with uniform optical index $n_0$. The previous set of equations enables the computation of the light field for all slices of the sample until the last slice of position $z_P$. The physical model is completed by these four elements: the position $z_D$ of the detection plane, the numerical aperture $NA$ of the optical system, the fact that the detector is sensitive to the modulus square of the light field and the background illumination $B_j$, the measurement in absence of the sample. Therefore, since $z_D<z_P$, the measurement $M_j$ is:
\[
\left\{
\begin{array}{l}
M_j=B_j.\left|   
FT^{-1}\left[FT\left[A^+_{P}\right].H^*_{Z_P-Z_D}.Aperture\right]
\right|^2\\
\text{with } Aperture(\vec\mu) = 
\left\{
\begin{array}{l}
1 \text{ if } |\vec\mu|\le \text{NA}/\lambda\\
0 \text{ if } |\vec\mu| > \text{NA}/\lambda\\
\end{array}\right.\\
\end{array}
\right.
\]
The above-mentioned forward model is able to predict $M_j(\mathcal R, \mathcal I)(\vec r)$ and the diffraction intensity measurements at point $\vec r$ of the detection plane for a sample with optical parameters $\mathcal R$ and $\mathcal I$ illuminated under incidence angle $j$.

The optimization criterion was set with the following Equation:
\[
\varepsilon(\mathcal R, \mathcal I)=\frac{1}{N_M}\sum_j \|\sqrt{M_j(\mathcal R, \mathcal I)}-\sqrt{m_j}\|_2^2+\frac{\alpha}{N_V}.TV(\mathcal R)+\frac{\beta}{N_V}.TV(\mathcal I)+\frac{\gamma}{N_V}.\|\mathcal R.(\mathcal R<0)\|_1
\]
where $N_M$ is the total number of measurements (number of image pixels $\times$ number of illumination angles) and $N_V$ is the number sample voxels. 
In the first term of the criterion, $\sqrt{m_j}$ was favored (compared to simply $m_j$) to create the data-fidelity term as it accounts for the Poisson nature of the detection noise. In the second and the third term, the operator $TV$ is the Total-Variation operator, which allows the denoising of the results while preserving the edges. The fourth term limits the values of $\mathcal R$ below $0$, as physically expected ($\mathcal R=\Delta RI>0$). $\alpha$, $\beta$ and $\gamma$ are weights to verify data-fidelity and regularity. 

Finally, $\mathcal R$ and $\mathcal I$ are obtained as the results of the optimization of $\varepsilon$ using the fast version of the iterative shrinkage-thresholding algorithms (FISTA) \cite{kamilov_optical_2016}. Error backpropagation is achieved with a PyTorch automatic differentiation feature. Reconstructions (40 iterations) last approximately 15 min for a 1024$\times$1024$\times$80 volume.

\begin{figure}[htbp]
    \centering
    \includegraphics[width=.6\textwidth]{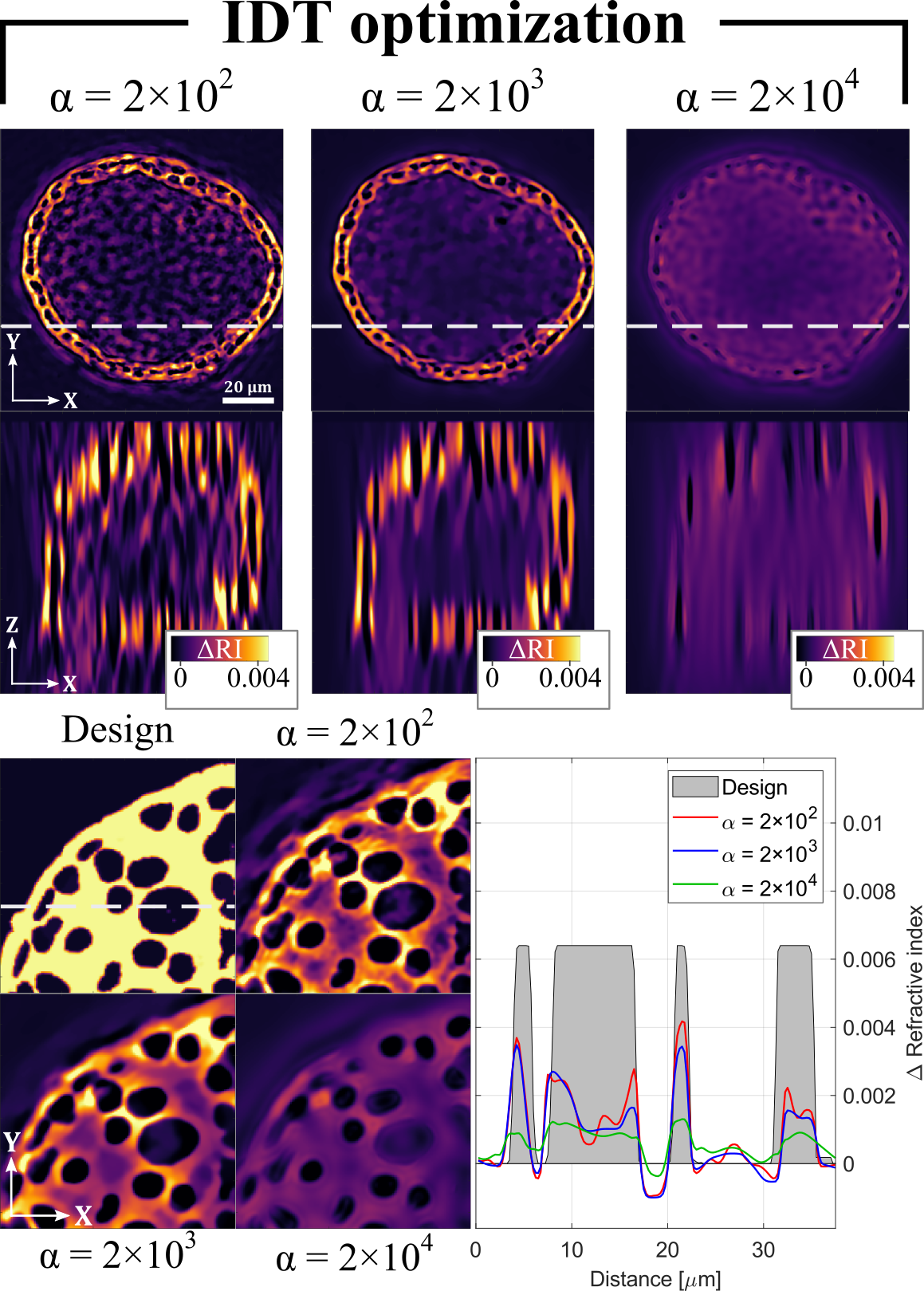} 
    \caption{Effect of different $\alpha$ on the IDT reconstructions of the healthy organoid at low RI contrast.}
    \label{fig:results_optimization}
\end{figure}

Reconstruction results are sensitive to the choice of weights $\alpha$, $\beta$ and $\gamma$; for example, in Fig.\ref{fig:results_optimization}, three results (under-regularized, acceptable and over-regularized) for $\mathcal R$ are presented for three values of parameter $\alpha$. 

\begin{figure}[htbp]
    \centering
    \includegraphics[width=.8\textwidth]{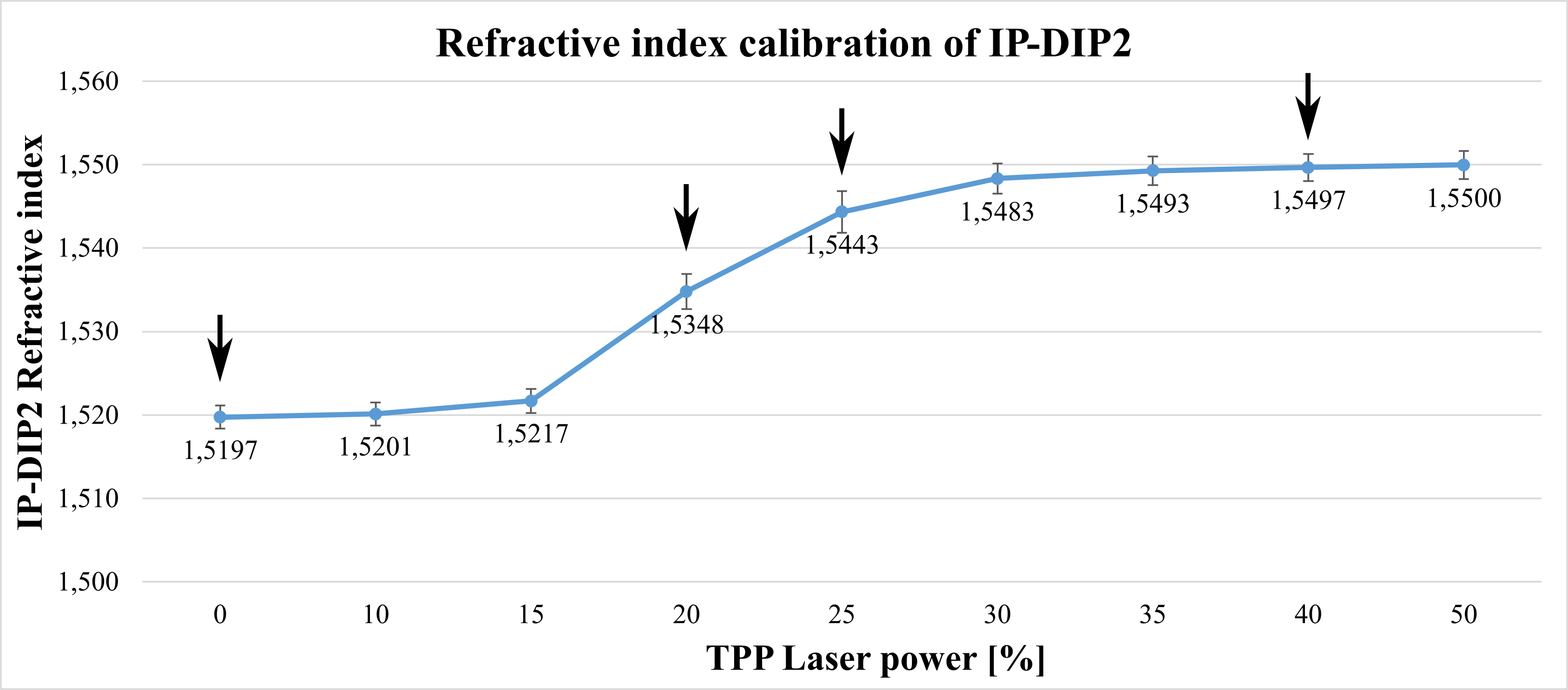} 
    \caption{RI calibration of the photopolymerized resin (IP-DIP2) at \SI{633}{\nano\meter} wavelength as a function of laser power setting during TPP fabrication. Arrows indicate particular RIs assigned to the regions of MLO and MTLO phantoms.}
    \label{fig:sup_ricalibration}
\end{figure}

\textbf{Refractive index calibration}
RI calibration curve used to assign the RI for various regions of the organoid phantoms was calculated as $\Delta RI = \lambda \Delta\phi/(2\pi \Delta d)$ ($\lambda$ -- \SI{633}{\nano\meter} illumination wavelength, $\Delta\phi$ -- phase shift of the wavefront introduced by the RI contrast and height difference $\Delta d$). For each laser power the results were averaged from four calibration step structures for a total of 36 steps (Fig.~\ref{fig:sup_ricalibration}) using the following key parameters of PPGT2 TPP system: 1.4 NA 63x microscope objective, galvo scan speed of \SI{10000}{\micro\meter/\second}, \SI{0.2}{\micro\meter} hatching distance, and \SI{0.3}{\micro\meter} slicing distance. RI between neighbouring data points has been linearly interpolated. Arrows indicate the RI assigned to the unpolymerized regions (e.g. lumen) and three RI values within the organoid phantoms. Average repeatability of the RI measured across four steps was 0,0002, while the error bars represent uncertainty of the RI considering the standard deviation of height and phase measurements, as well as the RI and temperature coefficients of the immersion liquid (Immersol 518F). Average RI uncertainty expanded with k=2 was found to be 0,0014.

Dispersion of the RI was a major consideration, since the measurements were performed at 532, 633 and 878 nm. The dispersion of the immersion oil and the polymer is similar (Cauchy parameters: B$_{imm} = 5.8561\cdot10^{-3}$ ${\mu}m^2$, C$_{imm} = 1.2990\cdot10^{-13}$ ${\mu}m^4$ according to the manufacturer's specification; B$_{resin} = 6.5456\cdot10^{-3}$ ${\mu}m^2$, C$_{resin} = 2.5345\cdot10^{-4}$ ${\mu}m^4$ \cite{Gissibl2017_ipdipRI}), therefore analyzing $\Delta$RI (relative to the RI of immersion for a given wavelength) compensates the offset of the RI. 
The upper bound of the dispersion contribution to the RI error has been estimated using the  maximal expected contrast at the extreme wavelengths: ${\Delta}RI_{max, 532} = 0.0400$ and ${\Delta}RI_{max, 878} = 0.0358$. Therefore, in the high-contrast case, the reconstructions performed at 532 nm are expected to have up to 12\% higher $\Delta$RI than the 878 nm. Since the experimental errors were much greater than that (in fact, IDT results at 532 nm were underestimated by a factor of 5.7 and 8.4 in terms of dry mass for the MLO and MTLO respectively when compared to the NIR-HT at 878 nm), we conclude that the RI dispersion error is not significant at this stage. Similarly, in the low-contrast case the dispersion between partially and fully cured resin is deemed to introduce even lower RI errors. Alternatively, the dispersion correction could be introduced to the data by segmenting regions encompassing one of the materials and multiplying its RI values by the appropriate factor. 

Measurements were performed at 24$\pm1^\circ$C. As in the case of dispersion, thermal coefficients of the immersion ($-0.00037/^\circ$C) and the polymer ($-0.00040/^\circ$C) are similar and thus the $\Delta$RI of the phantom is resilient to temperature fluctuations. Temperature difference of 3$^\circ$C yields $\Delta$RI $\approx$ 10$^{-4}$, which is currently below practical considerations.

\end{appendices}
\end{document}